\documentclass[12pt,a4paper]{article}
\usepackage[top=2.4cm,bottom=2.5cm,left=2.2cm,right=2.2cm]{geometry}
\usepackage{graphicx,latexsym,amsmath,amsfonts,amssymb,amsthm,amscd}

\newcommand{\be}{\begin{equation}}
\newcommand{\ee}{\end{equation}}
\renewcommand{\theequation}{\arabic{section}.\arabic{equation}}

\def\1{{\boldsymbol 1}}   %
\def\0{{\boldsymbol 0}}   %
\def\C{\mathbb{C}}        %
\def\N{\mathbb{N}}        %
\def\R{\mathbb{R}}        %
\def\cA{\mathcal{A}}      %
\def\cH{\mathcal{H}}      %
\def\cI{\mathcal{I}}      %
\def\cK{\mathcal{K}}      %
\def\cM{\mathcal{M}}      %
\def\cS{\mathcal{S}}      %
\def\fc{\mathfrak{c}}     %
\def\BC{{\rm BC}}         %
\def\diag{{\rm diag}}     %
\def\ri{{\rm i}}          %

\begin{document}
\rightline{}\vspace{1.0cm}

\begin{center}
{\large\bf
Equivalence of two sets of Hamiltonians associated with the rational $\boldsymbol{\BC_n}$ Ruijsenaars-Schneider-van Diejen system}
\end{center}

\vspace{0.2cm}

\begin{center}
T.F.~G\"orbe${}^a$ and L.~Feh\'er${}^{a,b}$\\

\bigskip
${}^a$Department of Theoretical Physics, University of Szeged\\
Tisza Lajos krt 84-86, H-6720 Szeged, Hungary\\
e-mail: tfgorbe@physx.u-szeged.hu

\medskip
${}^b$Department of Theoretical Physics, WIGNER RCP, RMKI \\
H-1525 Budapest, P.O.B.~49,  Hungary\\
e-mail: lfeher@physx.u-szeged.hu

\bigskip
\bigskip
\end{center}

\vspace{0.2cm}

\begin{abstract}
The equivalence of two  complete sets  of Poisson commuting Hamiltonians
of the (super)integrable rational $\BC_n$ Ruijsenaars-Schneider-van Diejen system
is established.
Specifically, the commuting Hamiltonians  constructed by van
Diejen are shown to be
linear combinations of the Hamiltonians generated by the characteristic polynomial
of the Lax matrix obtained recently by Pusztai, and the explicit formula of this
invertible linear transformation is found.
\end{abstract}

{\em Lax matrix}



\newpage

\section{Introduction}
\label{sec:1}
\setcounter{equation}{0}

Integrable many-body systems of Calogero-Moser type crop up in a wide range of physical
applications and are intimately related to important fields of mathematics \cite{OP1,RuijR,SuthR,PolR,EtiR}.
They occur in
rational, trigonometric/hyperbolic and elliptic families according to the functional form
of the Hamiltonian that inherently also involves a crystallographic root system. A further
significant feature is the existence of interesting deformations and extensions maintaining
integrability, as is exemplified by relativistic \cite{RS86} and spin Calogero-Moser
systems \cite{GH84}.
In this paper we shall deal with the $\BC_n$ generalization of the relativistic rational
Ruijsenaars-Schneider system, which is the simplest member of the systems
discovered by van Diejen \cite{vD1,vD2}.
We shall stay at the level of classical mechanics, where the $\BC_n$ rational `RSvD system'
is defined by the Hamiltonian\footnote{A deformation parameter $\beta>0$ can be introduced
by setting $H_\beta(\lambda, \theta) := H(\beta^{-1}\lambda,\beta\theta)$. Taking Taylor
expansion of $H_\beta$ in $\beta$, the leading term reproduces the usual $\BC_n$ rational
Calogero-Moser Hamiltonian.}
\begin{align}
H(\lambda,\theta)=&\sum_{j=1}^n\cosh(\theta_j)
\bigg[1+\frac{\nu^2}{\lambda_j^2}\bigg]^{\tfrac{1}{2}}
\bigg[1+\frac{\kappa^2}{\lambda_j^2}\bigg]^{\tfrac{1}{2}}
\prod_{\substack{k=1\\(k\neq j)}}^n
\bigg[1+\frac{\mu^2}{(\lambda_j-\lambda_k)^2}\bigg]^{\tfrac{1}{2}}
\bigg[1+\frac{\mu^2}{(\lambda_j+\lambda_k)^2}\bigg]^{\tfrac{1}{2}}\nonumber\\
&+\frac{\nu\kappa}{\mu^2}\prod_{j=1}^n
\bigg[1+\frac{\mu^2}{\lambda_j^2}\bigg]
-\frac{\nu\kappa}{\mu^2}.
\label{1.1}
\end{align}
Here $\mu,\nu,\kappa$ are real parameters for which we impose the
conditions $\mu\neq 0$, $\nu\neq 0$ and $\nu\kappa\geq 0$.
The generalized momenta $\theta=(\theta_1,\dots,\theta_n)$ run over $\R^n$ and
the `particle positions' $\lambda=(\lambda_1,\dots,\lambda_n)$ vary in
the Weyl chamber
\be
\fc=\{x\in\R^n\mid x_1>\dots>x_n>0\}.
\label{1.2}
\ee
In the work \cite{vD2,vD3} first the commutativity of $n$ quantum Hamiltonian difference operators was
proved.
It was then shown \cite{vD3,vD4} that the classical limit yields a Poisson commuting family having
the right functional rank for a Liouville integrable system.
Except for the rational case,
it is still an open problem to generate the classical Hamiltonians of van Diejen from a Lax matrix,
which would provide a useful tool for analyzing the dynamics of these systems.
A Lax matrix whose trace is the rational RSvD Hamiltonian \eqref{1.1} and whose higher spectral invariants
provide $n$ independent commuting Hamiltonians  was recently found by Pusztai \cite{P12}.
In the  papers \cite{P12,P13} the action-angle duality
between the hyperbolic $\BC_n$ Sutherland system and the rational $\BC_n$
RSvD system was also explored together with
the scattering properties of these systems.

The question we answer in this paper is the following.
What is the relationship between the commuting
Hamiltonians introduced by van Diejen and the ones generated by Pusztai's Lax matrix?
Both commuting families contain the `main Hamiltonian' \eqref{1.1} and exhibit rational dependence
 on  the
positions and exponential dependence on the momenta.
Thus one strongly expects that these two sets of commuting Hamiltonians can be expressed
in terms of each other.
Nevertheless, the question appears to be non-trivial since
 the Hamiltonian $H$ is maximally superintegrable \cite{AFG}, which entails that it is the member
 of several inequivalent  families of $n$ functionally independent functions in involution.

Here, we shall
 demonstrate that the $n$ Hamiltonians of van Diejen are linear combinations
 of the coefficients of the characteristic polynomial of the Lax matrix of \cite{P12}.
The transformation between the two sets will be shown to be invertible, and
its explicit form will be given as well.
Our arguments will rely on the action-angle map constructed with the help of
 Hamiltonian reduction in \cite{P12}. This is captured by a symplectomorphism
\be
\cS\colon\fc\times\R^n\to\fc\times\R^n,\quad
(q,p)\mapsto (\lambda,\theta),\qquad
\cS^\ast\bigg(\sum_{k=1}^nd\lambda_k\wedge d\theta_k\bigg)
=\sum_{k=1}^ndq_k\wedge d p_k,
\label{1.3}
\ee
such that $H\circ\cS$ depends only the action variables $q_k$.
(The notation fits the fact that the components of $q$
serve as position variables for the dual system.)
Although an explicit formula of the action-angle map is not available, we can compute
the action-angle transform of the commuting Hamiltonians of interest by utilizing that
 \cite{P13} the $H$-trajectory $(\lambda(t),\theta(t))$ with
initial condition $(\lambda,\theta)\in\fc\times\R^n$
has the $t\to\infty$ asymptotics
\be
\lambda_k(t)\sim t\sinh(q_k)-p_k
\quad\text{and}\quad
\theta_k(t)\sim q_k,
\qquad
k=1,\ldots,n,
\label{1.4}
\ee
with $(q,p)=\cS^{-1}(\lambda,\theta)$.
This will allow us to eventually show that the two sets of Hamiltonians at issue
correspond to two generating sets of the Weyl group invariant polynomials
in the variables $e^{\pm q_k}$ (restricted to the Weyl chamber $\fc$).
As was already mentioned, we shall also find the explicit relationship.
As a byproduct, we obtain an algebraic formula for the
characteristic polynomial of Pusztai's Lax matrix, which generalizes
well-known determinant identities for Cauchy-like matrices.

Section \ref{sec:2} describes the two families of Hamiltonians in play
and specifies how they share the main Hamiltonian $H$ \eqref{1.1}.
Our contribution is given by Proposition 1, Proposition 2  and Remark 3 in
Section \ref{sec:3}.
Section \ref{sec:4} offers a short discussion of the results and open problems.
There is also an appendix, where a useful formula of \cite{vD2} is presented.

\section{Two families of commuting Hamiltonians}
\label{sec:2}
\setcounter{equation}{0}

\subsection{Hamiltonians due to van Diejen}
\label{subsec:2.1}

In \cite{vD1,vD4} the following complete set of Poisson commuting Hamiltonians was given:
\be
H_l(\lambda,\theta)=\sum_{\substack{J\subset\{1,\ldots,n\},\ |J|\leq l\\
\varepsilon_j=\pm 1,\ j\in J}}
\cosh(\theta_{\varepsilon J})
V_{\varepsilon J;J^c}^{1/2}
V_{-\varepsilon J;J^c}^{1/2}
U_{J^c,l-|J|},\quad
l=1,\ldots,n,
\label{2.1}
\ee
with
\be
\begin{split}
\theta_{\varepsilon J}&=\sum_{j\in J}\varepsilon_j\theta_j,\\
V_{\varepsilon J;K}&=\prod_{j\in J}w(\varepsilon_j\lambda_j)
\prod_{\substack{j,j'\in J\\j<j'}}
v^2(\varepsilon_j\lambda_j+\varepsilon_{j'}\lambda_{j'})
\prod_{\substack{j\in J\\k\in K}}v(\varepsilon_j\lambda_j+\lambda_k)
v(\varepsilon_j\lambda_j-\lambda_k),\\
U_{K,p}&=(-1)^p\sum_{\substack{I\subset K,\ |I|=p\\\varepsilon_i=\pm 1,\ i\in I}}
\bigg(\prod_{i\in I}w(\varepsilon_i\lambda_i)\prod_{\substack{i,i'\in I\\i<i'}}
v(\varepsilon_i\lambda_i+\varepsilon_{i'}\lambda_{i'})
v(-\varepsilon_i\lambda_i-\varepsilon_{i'}\lambda_{i'})\\
&\hspace{10em}\times\prod_{\substack{i\in I\\k\in K\setminus I}}
v(\varepsilon_i\lambda_i+\lambda_k)v(\varepsilon_i\lambda_i-\lambda_k)\bigg).
\end{split}
\label{2.2}
\ee
It is worth noting that $J^c$ in (\ref{2.1}) denotes the complementary set,
and the contribution to $H_l$ coming from $J=\emptyset$ is $U_{\emptyset^c,l}$.
The relatively simple form of $U_{K,p}$ above was found in \cite{vD4}.
Equation \eqref{2.1} makes sense for $l=0$, as well, giving $H_0\equiv 1$.
In the rational case the functions $v$ and $w$ take the following
form\footnote{The parameters appearing in \cite{vD1,vD4} can be recovered
by introducing $\beta$ as in footnote 1 and then writing $\mu, \mu_0, \mu_0'$ for
$\beta\mu, \beta \nu, \beta \kappa$, respectively.
In the convention of
\cite{P13},
our $\mu$, $\theta$ and $q$  correspond to $2\mu$, $2\theta$ and $2q$.}
\be
v(x)=\frac{x+\ri\mu}{x},\quad
w(x)=\bigg[\frac{x+\ri\nu}{x}\bigg]
\bigg[\frac{x+\ri\kappa}{x}\bigg].
\label{2.3}
\ee
Up to irrelevant constants, $H_1$ reproduces
the Hamiltonian $H$ \eqref{1.1}. Indeed, one can check
that $H_1=2(H-n)$.

Take any $(\lambda,\theta)\in\fc\times\R^n$, set
$(q,p)=\cS^{-1}(\lambda,\theta)$ and consider the $H$-trajectory
$(\lambda(t),\theta(t))$ with initial condition $(\lambda,\theta)$.
Notice that the Hamiltonian $H_l$ \eqref{2.1} is constant along the $H$-trajectory.
By  utilizing the asymptotics \eqref{1.4},  one can readily check that
\be
(\cS^\ast H_l)(q,p)
=\lim_{t\to\infty}H_l(\lambda(t),\theta(t))
=\sum_{\substack{J\subset\{1,\ldots,n\},\ |J|\leq l\\
\varepsilon_j=\pm 1,\ j\in J}}
(-2)^{l-|J|}{n-|J|\choose l-|J|}\cosh(q_{\varepsilon J}).
\label{2.5}
\ee
From now on we let $\cH_l$ stand for the pullback $\cS^\ast H_l$  just computed,  and
stress that it depends only on the variable $q$.

\subsection{Hamiltonians obtained from the Lax matrix}
\label{subsec:2.2}

We recall some relevant objects of \cite{P12}.
First, prepare the $2n\times 2n$ Hermitian,
unitary matrix
\be
C=\begin{bmatrix}\0_n&\1_n\\\1_n&\0_n\end{bmatrix}
\label{2.6}
\ee
and the $2n\times 2n$ Hermitian matrix
\be
h(\lambda)=\begin{bmatrix}
a(\diag(\lambda))&b(\diag(\lambda))\\
-b(\diag(\lambda))&a(\diag(\lambda))
\end{bmatrix}
\label{2.7}
\ee
containing the smooth functions $a(x),b(x)$ given on the interval
$(0,\infty)\subset\R$ by
\be
a(x)=\frac{\sqrt{x+\sqrt{x^2+\kappa^2}}}{\sqrt{2x}},\quad
b(x)=\ri\kappa\frac{1}{\sqrt{2x}}\frac{1}{\sqrt{x+\sqrt{x^2+\kappa^2}}}.
\label{2.8}
\ee
Then introduce the vectors $z(\lambda)\in\C^n$, $F(\lambda,\theta)\in\C^{2n}$
by the formulae
\be
z_l(\lambda)=-\bigg[1+\frac{\ri\nu}{\lambda_l}\bigg]
\prod_{\substack{m=1\\(m\neq l)}}^n\bigg[1+\frac{\ri\mu}{\lambda_l-\lambda_m}\bigg]
\bigg[1+\frac{\ri\mu}{\lambda_l+\lambda_m}\bigg],
\label{2.9}
\ee
and
\be
F_l(\lambda,\theta)=e^{-\tfrac{\theta_l}{2}}|z_l(\lambda)|^{\tfrac{1}{2}},\quad
F_{n+l}(\lambda,\theta)=\overline{z_l(\lambda)}
F_l(\lambda,\theta)^{-1},
\label{2.10}
\ee
$l=1,\dots,n$.
With these notations at hand,
the $2n\times 2n$ matrix
\be
A_{j,k}(\lambda,\theta)=\frac{\ri\mu F_j\overline{F_k}
+\ri(\mu-2\nu)C_{j,k}}{\ri\mu+\Lambda_j-\Lambda_k},\quad
j,k\in\{1,\ldots,2n\},
\label{2.11}
\ee
with $\Lambda=\diag(\lambda,-\lambda)$ is used to define the `RSvD Lax matrix' \cite{P12}:
\be
L(\lambda,\theta)=h(\lambda)^{-1}A(\lambda,\theta)h(\lambda)^{-1}.
\label{2.12}
\ee
The matrices $h$, $A$, and $L$ are invertible and
satisfy the relations
\be
ChC=h^{-1},\quad
CAC=A^{-1},\quad
CLC=L^{-1}.
\label{2.13}
\ee
Their determinants are
\be
\det(h)=\det(A)=\det(L)=1.
\label{2.14}
\ee
Let $K_m$ denote the coefficients of the characteristic polynomial of
$L$ \eqref{2.12},
\be
\det(L(\lambda,\theta)-x\1_{2n})
=K_0(\lambda,\theta)x^{2n}+K_1(\lambda,\theta)x^{2n-1}+\dots
+K_{2n-1}(\lambda,\theta)x+K_{2n}(\lambda,\theta).
\label{2.15}
\ee
An immediate consequence of \eqref{2.13},\eqref{2.14} is that
\be
K_{2n-m}\equiv K_m,\quad m=0,1,\dots,n,
\label{2.16}
\ee
thus the functions $K_0\equiv 1,K_1,\dots,K_n$ fully determine the characteristic
polynomial \eqref{2.15}. The first non-constant member of this family is
proportional to $H$ \eqref{1.1}, that is $K_1=-2H$.
The asymptotic form of the Lax matrix $L$ \eqref{2.14} is the diagonal matrix
\be
\diag(e^{-q},e^{q}),
\label{2.17}
\ee
hence the action-angle transforms of the functions $K_m$ ($m=0,1,\dots,n$) can be easily
computed to be
\be
(\cS^\ast K_m)(q,p)=(-1)^m
\sum_{a=0}^{\big\lfloor\tfrac{m}{2}\big\rfloor}
\sum_{\substack{J\subset\{1,\ldots,n\},\ |J|=m-2a\\\varepsilon_j=\pm 1,\ j\in J}}
{n-|J|\choose a}\cosh(q_{\varepsilon J}).
\label{2.18}
\ee
Of course, we used the asymptotics  (\ref{1.4}) and that $K_m$ is constant along the flow of $H$.
Now we introduce the shorthand $\cK_m:= \cS^\ast K_m$, and observe that it only depends on $q$.

\section{Relation between the two families of Hamiltonians}
\label{sec:3}
\setcounter{equation}{0}

It is worth emphasizing that finding a formula relating the families
$\{H_l\}_{l=0}^n$ and $\{K_m\}_{m=0}^n$ is equivalent to finding a
relation between their action-angle transforms $\{\cH_l\}_{l=0}^n$
and $\{\cK_m\}_{m=0}^n$.

\bigskip\noindent\bf Proposition 1. \it
There exists an invertible linear relation between the two families
$\{\cH_l\}_{l=0}^n$ and $\{\cK_m\}_{m=0}^n$.\rm

\begin{proof}
Let us introduce the auxiliary functions
\be
\cM_k(q)=\sum_{\substack{J\subset\{1,\ldots,n\},\ |J|=k\\
\varepsilon_j=\pm 1,\ j\in J}}\cosh(q_{\varepsilon J}),
\quad q\in\R^n,\quad k=0,1,\dots,n.
\label{3.1}
\ee
For any $l\in\{0,1,\dots,n\}$ the Hamiltonian $\cH_l$ \eqref{2.5}
is a linear combination of $\cM_0,\cM_1,\dots,\cM_l$,
\be
\cH_l(q)=\sum_{k=0}^l(-2)^{l-k}{n-k\choose l-k}\cM_k(q).
\label{3.2}
\ee
This shows that the matrix of the linear map transforming $\{\cM_k\}_{k=0}^n$
into $\{\cH_l\}_{l=0}^n$ is lower triangular with ones on the diagonal,
hence the above relation is invertible.
Similarly, any function $\cK_m$ \eqref{2.18}, $m\in\{0,1,\dots,n\}$ can be
expressed as a linear combination of $\cM_m,\cM_{m-2},\dots,\cM_3,\cM_1$ or
$\cM_m,\cM_{m-2},\dots,\cM_2,\cM_0$ depending on the parity of $m$, that is
\be
\cK_m(q)=(-1)^m\sum_{a=0}^{\big\lfloor\tfrac{m}{2}\big\rfloor}
{n-(m-2a)\choose a}\cM_{m-2a}(q).
\label{3.3}
\ee
Hence the linear transformation relating $\{\cM_k\}_{k=0}^n$ to $\{\cK_m\}_{m=0}^n$
has a lower triangular matrix with diagonal components $\pm 1$, implying that it is
invertible. This proves the existence of an invertible linear relation between the
two families $\{\cH_l\}_{l=0}^n$ and $\{\cK_m\}_{m=0}^n$.
\end{proof}

Now, we prove an explicit formula expressing $\cH_l$ as linear combination of
$\{\cK_m\}_{m=0}^l$.

\medskip\noindent\bf Proposition 2. \it
For any fixed $n\in\N$, $l\in\{1,\ldots,n\}$ and $q\in\R^n$ we have
\be
(-1)^l\cH_l(q)=\cK_l(q)+
\sum_{m=0}^{l-1}\frac{2(n-m)}{2(n-m)-(l-m)}{(n-l)+(n-m)\choose l-m}\cK_m(q).
\label{3.4}
\ee\rm

\begin{proof}
Substitute $\cK_m$ \eqref{2.18} into the right-hand side of the expression
above to obtain
\begin{multline}
\sum_{k=0}^{l-1}\sum_{a=0}^{\big\lfloor\tfrac{k}{2}\big\rfloor}
\sum_{\substack{J\subset\{1,\ldots,n\},\ |J|=k-2a\\\varepsilon_j=\pm 1,\ j\in J}}
(-1)^k\frac{2(n-k)}{2(n-k)-(l-k)}{(n-l)+(n-k)\choose l-k}\times\\
\times{n-(k-2a)\choose a}\cosh(q_{\varepsilon J})
+\sum_{a=0}^{\big\lfloor\tfrac{l}{2}\big\rfloor}
\sum_{\substack{J\subset\{1,\ldots,n\},\ |J|=l-2a\\\varepsilon_j=\pm 1,\ j\in J}}
(-1)^l{n-(l-2a)\choose a}\cosh(q_{\varepsilon J}).
\label{3.5}
\end{multline}
Since $k=|J|+2a$ it is obvious that $(-1)^k=(-1)^{-|J|}$. Multiply \eqref{3.5} by
$(-1)^l$ and change the order of summations over $a$ and $J$ to get
\begin{multline}
\sum_{\substack{J\subset\{1,\ldots,n\},\ |J|<l\\\varepsilon_j=\pm 1,\ j\in J}}
(-1)^{l-|J|}\sum_{a=0}^{\big\lfloor\tfrac{l-|J|}{2}\big\rfloor}
\frac{2[n-(|J|+2a)]}{2[n-(|J|+2a)]-[l-(|J|+2a)]}\times\\
\times{(n-l)+(n-(|J|+2a))\choose l-(|J|+2a)}
{n-|J|\choose a}\cosh(q_{\varepsilon J})
+\sum_{\substack{J\subset\{1,\ldots,n\},\ |J|=l\\\varepsilon_j=\pm 1,\ j\in J}}
\cosh(q_{\varepsilon J}).
\label{3.6}
\end{multline}
Now, comparison of \eqref{3.2} with \eqref{3.6} leads to a relation equivalent to
\eqref{3.4},
\begin{multline}
\sum_{a=0}^{\big\lfloor\tfrac{l-|J|}{2}\big\rfloor}
\frac{2[n-(|J|+2a)]}{2[n-(|J|+2a)]-[l-(|J|+2a)]}\times\\
\times{2n-(l+|J|+2a)\choose l-(|J|+2a)}
{n-|J|\choose a}\bigg/{n-|J|\choose l-|J|}=2^{l-|J|}.
\label{3.7}
\end{multline}
For $n=l$ in \eqref{3.7} one obtains
\be
\begin{cases}\displaystyle
2\sum_{a=0}^{\big\lfloor\tfrac{l-|J|}{2}\big\rfloor}{l-|J|\choose a}=2^{l-|J|},&
\text{if}\ l-|J|\ \text{is odd,}\\[12pt]
\displaystyle
2\sum_{a=0}^{\tfrac{l-|J|}{2}-1}{l-|J|\choose a}
+{l-|J|\choose \frac{l-|J|}{2}}=2^{l-|J|},&
\text{if}\ l-|J|\ \text{is even,}\\
\end{cases}
\label{3.8}
\ee
which are well-known identities for the binomial coefficients.
This means that \eqref{3.4} holds for $l=n$ for all $n\in\N$,
which implies that if we consider $n+1$ variables it is sufficient
to check the cases $l<n+1$. With that in mind let us progress by induction
on $n$ and suppose that \eqref{3.4} is verified for all $1\leq l\leq n$
for some $n\in\N$.

First, notice that the Hamiltonians $\cH_l$ \eqref{2.5} satisfy
the following recursion
\be
\cH_l(q_1,\dots,q_n,q_{n+1})
=\cH_l(q_1,\dots,q_n)+4\sinh^2(\frac{q_{n+1}}{2})\cH_{l-1}(q_1,\dots,q_n).
\label{3.9}
\ee
This can be checked either directly or by utilizing that $\cH_l$ is
the $l$-th elementary symmetric function with variables $\sinh^2(\frac{q_i}{2})$
(see Appendix A).
Similarly, the functions $\cK_k$ \eqref{2.18} satisfy
\be
\cK_k(q_1,\dots,q_n,q_{n+1})
=\cK_k(q_1,\dots,q_n)
-2\cosh(q_{n+1})\cK_{k-1}(q_1,\dots,q_n)
+\cK_{k-2}(q_1,\dots,q_n),
\label{3.10}
\ee
with $\cK_{-1}\equiv 0$. Let us introduce some shorthand notation,
such as the $\R^{l+1}$ vectors
\be
\vec\cH(n):=(\cH_0,-\cH_1,\dots,(-1)^l\cH_l)^\top\quad\text{and}\quad
\vec\cK(n):=(\cK_0,\cK_1,\dots,\cK_l)^\top
\label{3.11}
\ee
and the $\R^{(l+1)\times(l+1)}$ matrices
\be
\cA(n)_{j+1,k+1}:=\begin{cases}\displaystyle
\frac{2(n-k)}{2(n-k)-(j-k)}{(n-j)+(n-k)\choose j-k},
&\text{if}\ j\geq k,\\[1em]
0,&\text{if}\ j<k,
\end{cases}
\label{3.12}
\ee
where $j,k\in\{0,\ldots,l\}$ and
\be
\cH(n,n+1):=\1_{l+1}-4\sinh^2(\frac{q_{n+1}}{2})\cI_{-1},
\quad
\cK(n,n+1):=\1_{l+1}-2\cosh(q_{n+1})\cI_{-1}+\cI_{-2}
\label{3.13}
\ee
with $(\cI_{-m})_{j+1,k+1}:=\delta_{j,k+m}$, $m>0$.
The relations \eqref{3.9} and \eqref{3.10} can be written in the concise form
\be
\vec\cH(n+1)=\cH(n,n+1)\vec\cH(n),\quad
\vec\cK(n+1)=\cK(n,n+1)\vec\cK(n)
\label{3.14}
\ee
and our assumption is condensed into
\be
\vec\cH(n)=\cA(n)\vec\cK(n).
\label{3.15}
\ee
Using this notation it is clear that the desired
induction step is equivalent to
the matrix equation
\be
\cH(n,n+1)\cA(n)=\cA(n+1)\cK(n,n+1).
\label{3.16}
\ee
Spelling this out at some arbitrary $(j,k)$-th entry gives us
\begin{multline}
\frac{A+B}{A}{A\choose B}
-4\sinh^2\bigg(\frac{\alpha}{2}\bigg)\frac{A+B}{A+1}{A+1\choose B-1}
=\\[.5em]
=\frac{A+B+2}{A+2}{A+2\choose B}
-2\cosh(\alpha)\frac{A+B}{A+1}{A+1\choose B-1}
+\frac{A+B-2}{A}{A\choose B-2},
\label{3.17}
\end{multline}
where
\be
A:=2n-j-k,\quad B:=j-k,\quad\alpha:=q_{n+1}.
\label{3.18}
\ee
A simple direct calculation shows that \eqref{3.17} indeed holds implying that
\eqref{3.4} is also true for $n+1$ for any $l\leq n$. The case $l=n+1$ is given
by the argument preceding induction. This completes the proof.
\end{proof}

\medskip\noindent{\bf Remark 3.}
We showed in Proposition 1 that the relation  \eqref{3.4} is invertible.
Without spending space on the proof, we note that the inverse relation can be written explicitly as
\be
(-1)^m\cK_m(q)=\sum_{l=0}^{m}{2(n-l)\choose m-l}\cH_l(q).
\label{3.19}
\ee
\rm

\section{Discussion}
\label{sec:4}
\setcounter{equation}{0}

In this paper we demonstrated that
the commuting
Hamiltonians of the rational RSvD system constructed originally by van Diejen
are linear combinations of the coefficients of the characteristic polynomial of the
Lax matrix found recently by Pusztai, and vice versa.
The derivation utilized the action-angle map and the scattering theory
results of \cite{P12,P13}.
Our Proposition 2 gives rise to a determinant representation of the
somewhat complicated expressions $H_l$ in (\ref{2.1}).
It could be of some interest to provide a purely algebraic proof of the
resulting formula of the characteristic polynomial of the Lax matrix.

The configuration space $\fc$ (\ref{1.2}) is an open Weyl chamber
associated with the  Weyl group $W(\BC_n)$,  and
after extending this domain all Hamiltonians
that we dealt with enjoy $W(\BC_n)$ invariance. In particular, the sets $\{\cH_l\}_{l=0}^n$,
$\{ \cK_l\}_{l=0}^n$ and $\{\cM_l\}_{l=0}^n$ represent different free generating sets
of the  invariant polynomials in the functions $e^{\pm q_k}$ ($k=1,\ldots, n$) of
the action variables $q_k$ acted upon by the sign changes and permutations that form
$W(\BC_n)$. In order to verify this, it is useful to point out that the $W(\BC_n)$
invariant polynomials in the variables $e^{\pm q_k}$ are the same as the ordinary
symmetric polynomials in the variables $\cosh(q_k)$.
The statement that $\{\cH_l\}_{l=0}^n$ is a free generating set for these polynomials
then follows, for example,
from the identity presented in Appendix A.

Analogous statements hold obviously also for the different real form of the complex
rational RSvD system studied in \cite{FG14}, which is also superintegrable.

An interesting open problem for future work is to extend the considerations reported here
to the hyperbolic RSvD system having five independent coupling parameters.

\bigskip\bigskip \noindent\bf Acknowledgements. \rm
We are greatly indebted to
J.F.~van Diejen for suggesting us to use the asymptotics of the $H$-flow
in order to prove the formula of Proposition 2, which we originally found and tried to prove in
a rather roundabout way. L.F. also wishes to thank S. Ruijsenaars for helpful discussions.
This work was supported in part by the Hungarian Scientific Research
Fund (OTKA) under the grant K-111697.

\appendix
\renewcommand{\theequation}{\Alph{section}.\arabic{equation}}

\section{$\cH_l$ as elementary symmetric function}

Fix an arbitrary $n\in\N$ and $l\in\{0,1,\ldots,n\}$ and let $e_l$ stand for the
$l$-th elementary symmetric polynomial in $n$ variables $x_1,\ldots,x_n$, i.e., $e_0(x_1,\dots,x_n)=1$ and for $l\geq 1$
\be
e_l(x_1,\ldots,x_n)=\sum_{1\leq j_1<\dots<j_l\leq n}x_{j_1}\cdots x_{j_l}.
\label{A.1}
\ee
In the text, we referred to the following useful result due to van Diejen
(\cite{vD2} Proposition 2.3). For convenience, we present it together
with a direct proof.

\medskip\noindent\bf Proposition A. \it
By using \eqref{2.5} it can be shown that
\be
\cH_l(q)=4^le_l(\sinh^2\frac{q_1}{2},\dots,\sinh^2\frac{q_n}{2}).
\label{A.2}
\ee\rm

\begin{proof}
First, $e_l$ has the equivalent form
\be
e_l(\sinh^2\frac{q_1}{2},\dots,\sinh^2\frac{q_n}{2})
=\sum_{J\subset\{1,\dots,n\},\ |J|=l}\;\prod_{j\in J}\sinh^2\frac{q_j}{2}.
\label{A.3}
\ee
Utilizing the identity $\sinh^2(\alpha/2)=[\cosh(\alpha)-1]/2$
casts the right-hand side into
\be
\sum_{J\subset\{1,\dots,n\},\ |J|=l}2^{-l}
\prod_{j\in J}[\cosh(q_j)-1]=
\sum_{J\subset\{1,\dots,n\},\ |J|=l}2^{-l}
\sum_{K\subset J}(-1)^{l-|K|}\prod_{k\in K}\cosh(q_k).
\label{A.4}
\ee
The two sums on the right-hand side can be merged into one,
but the multiplicity of subsets must remain the same.
This results in the appearance of a binomial coefficient
\begin{multline}
\sum_{J\subset\{1,\dots,n\},\ |J|\leq l}\frac{(-1)^{l-|J|}}{2^l}{n-|J|\choose l-|J|}
\prod_{j\in J}\cosh(q_j)=\\
=\sum_{\substack{J\subset\{1,\dots,n\},\ |J|\leq l\\\varepsilon_j=\pm 1,\ j\in J}}
\frac{(-1)^{l-|J|}}{2^{l+|J|}}{n-|J|\choose l-|J|}
\prod_{j\in J}\cosh(\varepsilon_j q_j),
\label{A.5}
\end{multline}
where we also used that $\cosh$ is an even function and compensated the
`over-counting' of terms. Now, let us simply pull a $4^{-l}$ factor out of the sum to get
\be
4^{-l}
\sum_{\substack{J\subset\{1,\dots,n\},\ |J|\leq l\\\varepsilon_j=\pm 1,\ j\in J}}
(-2)^{l-|J|}{n-|J|\choose l-|J|}\prod_{j\in J}\cosh(\varepsilon_j q_j).
\label{A.6}
\ee
Recall the following identity for the hyperbolic cosine of the sum of a finite number, say $N$, real arguments (see \cite{H1841} Art. 132 and apply $\cos(\ri\alpha)=\cosh(\alpha)$)
\be
\cosh\bigg(\sum_{k=1}^N\alpha_k\bigg)
=\bigg[\prod_{k=1}^N\cosh(\alpha_k)\bigg]
\bigg[\sum_{m=0}^{\big\lfloor\tfrac{ N}{2}\big\rfloor}e_{2m}(\tanh(\alpha_1),\dots,\tanh(\alpha_N))\bigg],
\label{A.7}
\ee
where $e_{2m}$ are now elementary symmetric functions with arguments
$\tanh(\alpha_1),\dots,\tanh(\alpha_N)$.
Note that for any $m>0$ and set of signs $\varepsilon$ there is another one
$\varepsilon'$, such that $e_{2m}^{J,\varepsilon'}=-e_{2m}^{J,\varepsilon}$,
therefore by using \eqref{A.7} we see that \eqref{A.6} equals to
\begin{multline}
4^{-l}
\sum_{\substack{J\subset\{1,\dots,n\},\ |J|\leq l\\\varepsilon_j=\pm 1,\ j\in J}}
(-2)^{l-|J|}{n-|J|\choose l-|J|}
\prod_{j\in J}
\cosh(\varepsilon_j q_j)\sum_{m=0}^{\big\lfloor\tfrac{|J|}{2}\big\rfloor}
s_{2m}^{J,\varepsilon}=\\
=4^{-l}
\sum_{\substack{J\subset\{1,\dots,n\},\ |J|\leq l\\\varepsilon_j=\pm 1,\ j\in J}}
(-2)^{l-|J|}{n-|J|\choose l-|J|}
\cosh(q_{\varepsilon J}).
\label{A.8}
\end{multline}
Applying \eqref{2.5} concludes the proof.
\end{proof}

\end{document}